\documentclass[11pt]{article}

\usepackage[a4paper, margin=1in]{geometry}

\usepackage{iftex}
\ifPDFTeX
  \usepackage[T1]{fontenc}
  \usepackage[utf8]{inputenc}
  \usepackage{amsthm}
  \usepackage{newtxtext}
  \usepackage{newtxmath}
  \usepackage[scaled=0.92]{helvet}
  \usepackage[light]{roboto}
  \usepackage{inconsolata}
  \newcommand{\dotstts}{\textsf{dots.tts}}
\else
  \usepackage{fontspec}
  \usepackage{xeCJK}
  \setCJKsansfont{Noto Sans CJK SC}
  \setCJKmonofont{Noto Sans Mono CJK SC}
  \newfontfamily\dotsttsfont[
    UprightFont    = *,
    ItalicFont     = *,
    BoldFont       = *,
    BoldItalicFont = *,
  ]{Roboto Slab}
  \newcommand{\dotstts}{{\dotsttsfont\upshape dots.tts}}
\fi
\usepackage{microtype}
\usepackage{setspace}
\usepackage[skip=6pt plus 1pt, indent=0pt]{parskip}

\usepackage{mathtools}
\usepackage{amsthm}
\ifPDFTeX\else
  \usepackage{amssymb}
\fi

\usepackage{graphicx}
\graphicspath{{figures/}}
\usepackage{booktabs}
\usepackage{multirow}
\usepackage{array}
\usepackage{tabularx}
\usepackage{makecell}
\usepackage{tablefootnote}
\usepackage{enumitem}
\setlist[itemize]{leftmargin=*, topsep=2pt, itemsep=1pt}
\setlist[enumerate]{leftmargin=*, topsep=2pt, itemsep=1pt}

\usepackage[font=small, labelfont=bf, skip=4pt]{caption}
\usepackage{subcaption}
\usepackage[table]{xcolor}
\usepackage{colortbl}
\usepackage[normalem]{ulem}

\definecolor{linkblue}{HTML}{1F4E79}
\definecolor{rowhl}{gray}{0.90}
\usepackage[colorlinks=true,
            linkcolor=linkblue,
            citecolor=linkblue,
            urlcolor=linkblue,
            breaklinks=true]{hyperref}
\usepackage[capitalise, noabbrev]{cleveref}

\usepackage[round, authoryear, sort]{natbib}
\usepackage{fancyhdr}
\makeatletter
\renewenvironment{abstract}{%
    \begin{center}%
        {\Large\bfseries \abstractname}%
    \end{center}%
    \quotation
}{%
    \endquotation
}
\makeatother

\title{\vspace{-1.0cm}\huge\bfseries \dotstts{} Technical Report}
\author{\textbf{\dotstts{} Team}}
\date{}

\begin{document}

\maketitle

\begin{center}
\small
\begin{tabular}{rl}
\textbf{Code:}        & \url{https://github.com/studio-dots-ai/dots.tts} \\
\textbf{Model:}       & \url{https://huggingface.co/collections/dots-studio/dotstts} \\
\textbf{Demo:}        & \url{https://studio-dots-ai.github.io/dots.tts-demo/} \\
\end{tabular}
\end{center}
\vspace{0.5em}

\begin{abstract}
\noindent We present \dotstts{}, a 2B-parameter continuous autoregressive
text-to-speech (TTS) foundation model that models speech in a
continuous latent space. Compared with existing continuous autoregressive models, our key innovations are threefold. First, we train
an AudioVAE with multiple objectives to build a semantically structured and
prediction-friendly continuous speech space. Second, we use full-history
conditioning in the flow-matching head to preserve long-range consistency and
reduce drift during generation. Third, we apply reward-free self-corrective
post-training to the flow-matching head to further improve robustness and acoustic
quality. After being trained on a large-scale multilingual corpus,
\dotstts{} achieves the best average performance on
Seed-TTS-Eval, with WERs of 0.94\%/1.30\%/6.60\% and SIM scores of
81.0/77.1/79.5 on the zh/en/zh-hard test sets, respectively.
Across other benchmarks, \dotstts{} also consistently demonstrates
open-source state-of-the-art performance, exhibiting strong generation
stability, voice cloning ability, and emotional expressiveness. For efficient inference, we further apply CFG-aware MeanFlow distillation,
enabling low-latency speech generation with first-packet latencies
of 85/54~ms in output streaming and dual-streaming modes, respectively. To facilitate reproducible research and practical deployment, we release the
training and inference code, together with the pretrained, post-trained, and
MeanFlow-distilled checkpoints, under the Apache~2.0 license.
\end{abstract}

\section{Introduction}
\label{sec:intro}

Text-to-speech (TTS) systems have largely solved intelligibility on
standard read-speech benchmarks. What users expect from a modern system
is broader: expressive and controllable output, real-time synthesis,
and coverage of neutral reading, emotional dialogue, paralinguistic
events, singing, and general audio. Current systems pursue this goal
along three roughly distinct technical routes, and each route has its
own unresolved problem.

The first route is non-autoregressive (NAR) generation on top of a
flow-matching~\citep{flowmatching} or Diffusion
infilling. Voicebox~\citep{voicebox}, F5-TTS~\citep{f5tts},
OmniVoice~\citep{omnivoice}, and
LongCat-AudioDiT~\citep{longcataudiodit} fall into this group. They
generate an utterance in a single parallel pass, and few-step
distillation~\citep{rectifiedflow, meanflow} can
push inference latency very low. NAR is a good fit for offline data synthesis, dubbing, and asset production,
whereas autoregressive (AR) generation is more natural for interactive use.
We focus on the AR side in this report.

Following the success of large language models, AR generation over
discrete tokens has become the mainstream production paradigm.
The CosyVoice family~\citep{cosyvoice3},
Qwen3-TTS~\citep{qwen3tts}, Llasa~\citep{llasa},
IndexTTS~\citep{indextts, indextts2}, and
Seed-TTS~\citep{seedtts} all follow some variant of this recipe.
Reducing speech to a discrete vocabulary lets
the system reuse most of the textual large language model (LLM) stack: stable next-token
prediction training, mature SFT and preference-optimization tooling, well-studied
scaling behavior, and efficient inference kernels. Hence its current
dominance in deployment.
The limits of discrete AR come from the tokenizer, it caps what the language model (LM) can
express. This bottleneck explains why discrete-token systems
struggle to cover speech, emotional paralinguistics, singing, and
ambient sound within a single distribution.

A recent line of work removes this bottleneck by modeling speech as
continuous-representation AR generation. These methods differ mainly in
their per-step distribution parameterization. KALL-E~\citep{kalle} predicts 
a Flow-VAE next-frame latent with a Kullback-Leibler divergence (KL) objective and no diffusion component.
DiTAR~\citep{ditar}, VibeVoice~\citep{vibevoice} and VoxCPM~\citep{voxcpm} pair
an LM backbone with a per-patch diffusion or flow-matching head, whereas
ARDiT~\citep{ardit} folds both roles into a decoder-only diffusion transformer.
Any2Speech~\citep{any2speech} argues more broadly for native-agentic generation
over continuous tokenizers. Avoiding discrete quantization raises the perceptual ceiling,
permits a single distribution over speech, paralinguistics, singing, and
general audio, and preserves the streaming, prompt-conditioned behavior
that makes LM-based systems easy to deploy.

The main issue which has so far kept this paradigm short of production
maturity is \emph{long-range error accumulation}. With
discrete tokens, the codec snaps an imperfect sample back to a valid
acoustic configuration before it reaches the waveform. Continuous
latents have no such quantization buffer: every small prediction error
is reconstructed faithfully by the decoder and then fed back as
conditioning for the next AR step.


\dotstts{} is our attempt to close the gap: an end-to-end AR TTS model
over continuous latents. The system combines three practices. First, we build
the system on a high-fidelity semantic AudioVAE, following the HoliTok-style
VAE recipe~\citep{holitok}. After reconstruction-oriented training, we
add multitask downstream objectives and a WavLM~\citep{wavlm}
representation-alignment loss, making the latent space both
semantic and learnable for the AR backbone. Second, inspired by ARDiT's
autoregressive diffusion formulation, we decompose continuous
generation into three specialized modules: a semantic encoder for
content-aligned representation, an LLM for long-range
text-to-content modeling, and a full-context AR flow-matching head for
local acoustic rendering. This separation keeps
semantic reasoning and acoustic rendering from competing inside a single
module, reducing error accumulation during long rollouts. Third, for
post-training, we adapt the reward-free self-correction idea of
SOAR~\citep{soar} to the AR flow-matching head, exposing the acoustic DiT
to its own off-trajectory inference errors without requiring a reward
model or external teacher. Together, these choices improve long-range
stability while preserving the fidelity and expressiveness enabled by
continuous-latent generation.

Trained on 1.5M hours of speech, \dotstts{} achieves state-of-the-art
stability and zero-shot voice-cloning quality on
Seed-\allowbreak{}TTS-\allowbreak{}Eval~\citep{seedttseval}, leads
average speaker similarity on the 24-language MiniMax multilingual
test set~\citep{minimaxspeech}, and is competitive on
EmergentTTS-\allowbreak{}Eval~\citep{emergentttseval} and
CV3-\allowbreak{}Eval~\citep{cosyvoice3}, which together cover stability,
speaker similarity, naturalness, prosody, paralinguistics, and
multilingual coverage.

For real-time and conversational use without
quality loss, the full model is designed from the outset to be causal at
the latent-patch level. This causal structure lets text-side planning
and audio-latent emission proceed incrementally, making 1T1A interleaved
dual-stream inference possible. We further apply MeanFlow
distillation to compress the flow-matching ODE to as few
as 2 to 4 function evaluations, yielding low first-packet latency on the
same backbone: 85~ms at RTF~0.231 in plain mode and 54~ms at RTF~0.245
in interleaved streaming mode. The full
efficiency profile, including voice-cloning scenarios and an audio-prompt
cache, is reported in \Cref{sec:efficiency}. Alongside the model, we
document a full training recipe for continuous AR TTS, covering
pretraining, our Self-corrective alignment stage for the flow-matching
head, and the MeanFlow distillation above.

In summary, our contributions are:

\begin{itemize}
    \item We present \dotstts{}, a 2B-parameter fully continuous
    end-to-end AR TTS system that removes discrete acoustic tokens while
    achieving state-of-the-art stability and zero-shot voice-cloning
    quality on Seed-TTS-Eval.
    \item We alleviate the instability of continuous AR generation with
    three complementary designs: a semantic causal AudioVAE, an
    ARDiT-inspired decomposition into semantic planning and acoustic
    rendering, and reward-free Self-corrective alignment for the
    flow-matching head.
    \item We improve inference efficiency with CFG-aware MeanFlow
    distillation and fully causal 1T1A interleaved streaming, enabling
    low first-packet latency (85~ms at RTF~0.231 plain, 54~ms at
    RTF~0.245 interleaved; \Cref{sec:efficiency}) suitable for real-time
    deployment.
\end{itemize}

\section{Model}
\label{sec:model}

\subsection{Overview}
\label{sec:model:overview}

\dotstts{} is a fully continuous, end-to-end autoregressive TTS system
built from two decoupled networks: an audio variational autoencoder
(AudioVAE) that defines the continuous representation, and an
autoregressive backbone that predicts that representation one patch at
a time. The AudioVAE encodes 48~kHz
mono speech into a 128-dimensional latent stream at 25~Hz ($1920\times$
temporal downsampling) and decodes it back via a BigVGAN-style
decoder~\citep{bigvgan}; once trained, it is frozen and serves as both
the generation target and the input representation for the backbone.

\begin{figure}[t]
    \centering
    \includegraphics[width=0.9\linewidth]{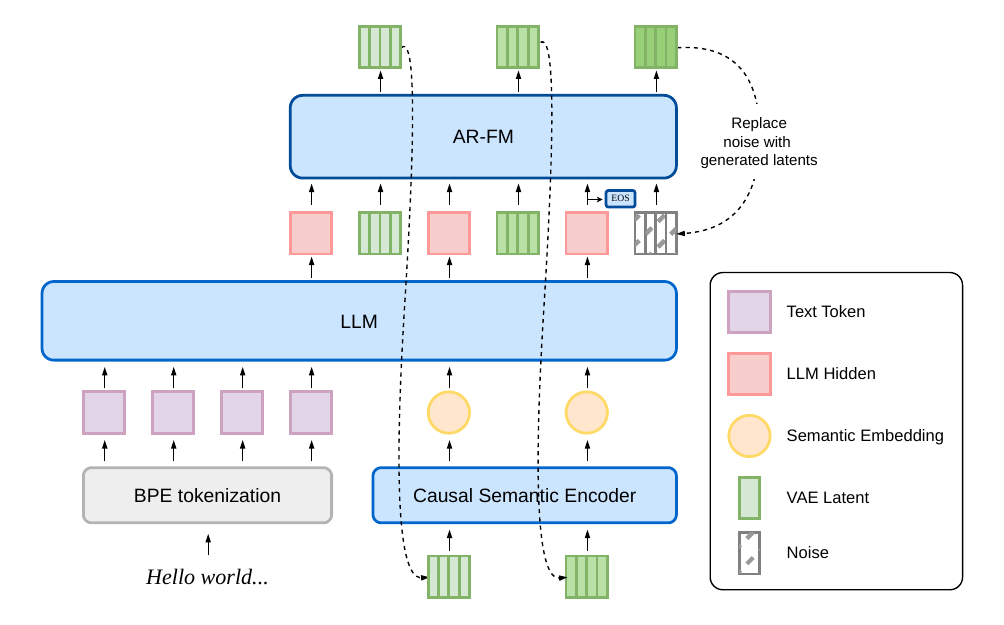}
    \caption{Overview of the \dotstts{} backbone. BPE text tokens and
    6.25~Hz audio-semantic embeddings share a single LLM stream; each
    LLM hidden state conditions the AR-FM head to generate the next
    four-frame VAE-latent patch, which is fed back through the semantic
    encoder at the next step. The AudioVAE
    (\Cref{sec:model:vae}) is trained separately and frozen here.}
    \label{fig:overview}
\end{figure}

The backbone has three components (\Cref{fig:overview}): a semantic
encoder, an LLM, and an autoregressive flow-matching head. The LLM
handles the semantic side of generation and the flow-matching head
handles the acoustic side. The LLM, initialized from a pretrained text
LLM, consumes BPE text tokens together with a 6.25~Hz audio-semantic
embedding stream and emits one hidden state per audio step. The text is
placed as a prefix in plain TTS mode, or interleaved with audio in a
1T1A layout for low-latency streaming
(\Cref{sec:model:llm}). The autoregressive flow-matching head (AR-FM)
conditions on that hidden state and generates the next four-frame
patch of the 25~Hz VAE latent. The semantic encoder, re-used from the
AudioVAE training pipeline, projects each newly generated patch back
into a single 6.25~Hz embedding that feeds the LLM at the next step.
The LLM sees only this semantic summary, not the raw VAE latent. We
found this necessary to keep continuous-AR rollouts stable.


\subsection{AudioVAE}
\label{sec:model:vae}

The AudioVAE follows and adapts the training recipe and model architecture
of HoliTok, with a causal variant of the BigVGAN-v2 decoder on the synthesis
side. The encoder is a fully causal convolutional
stack of strided residual blocks with downsample strides
$[2, 2, 2, 4, 6, 10]$, ending in a posterior projection that produces
per-frame mean and log-variance. Following KALL-E, we add
flow regularization on top of the standard KL prior to shape a smooth,
low-noise latent space. All convolutions on both sides are causal,
keeping the VAE compatible with strict streaming synthesis.

We train the AudioVAE in two stages. The first stage targets
reconstruction quality alone: following the BigVGAN-v2 recipe, we
combine multi-period plus multi-scale sub-band CQT adversarial loss,
multi-scale mel-spectral reconstruction loss, and feature-matching loss,
together with the KL\,+\,flow regularization above.

The second stage targets \emph{learnability}. Such a heavily
compressed latent is reconstructive but retains so much acoustic
variation that a downstream LLM struggles to use it as a generation
target. To make the space semantically structured without sacrificing
reconstruction, we keep the first-stage losses and add two supervisions
on the latent: (i) a frame-level alignment loss against a frozen WavLM
teacher, and (ii) a multitask downstream block consisting
of a small encoder followed by a small LLM head, jointly trained on a
mixture of ASR, emotion, and speaker classification objectives. The
downstream LLM is discarded afterwards; we retain only the encoder,
which the backbone reuses as its semantic frontend
(\Cref{sec:model:semantic}). This second stage improves
the downstream diffusion learnability of the latent while
preserving reconstruction quality; see \Cref{tab:audiovae} for the
full reconstruction comparison against discrete codecs and other
continuous representations.

\subsection{LLM backbone}
\label{sec:model:llm}

We initialize the LLM from Qwen2.5-1.5B Base~\citep{qwen25} and
feed it text directly as BPE instead of phonemes.
On the audio side it runs at 6.25~Hz: the semantic encoder
(\Cref{sec:model:semantic}) compresses each 25~Hz VAE latent patch into
one audio-semantic embedding for the next LLM step, and the AR-FM head
(\Cref{sec:model:arfm}) consumes the LLM hidden state to emit the next
four-frame VAE patch.

Training supports two sequence layouts: a \emph{plain mode}, in which
the full text is placed as a prefix before the audio span and used for
standard TTS, and a 1T1A interleaved mode for
low-latency dual streaming, detailed in \Cref{sec:model:llm:interleave}.
Only the sequence layout differs between the two, and the per-step
interface between LLM, semantic encoder, and AR-FM head is identical.
Text positions carry no loss. The LLM is optimized end-to-end through
the flow-matching gradient that propagates back from the AR-FM head.

Starting from a text LLM gives us better prosody, better text
normalization, and the ability to follow natural-language style
prompts. The cost is data: a BPE-driven backbone needs substantially
more speech-text pairs than a phoneme one. We scale the training corpus accordingly.

\subsubsection{Interleaved sequence for dual streaming mode}
\label{sec:model:llm:interleave}

In 1T1A interleaved mode, a single BPE text token alternates
with one 6.25~Hz audio step until a text-end marker, after which audio
continues alone:
\begin{equation*}
    \underbrace{T\ A\ T\ A\ \ldots\ T\ A}_{\text{interleaved span}}\
    \langle\textsc{eot}\rangle\
    \underbrace{A\ A\ \ldots\ A}_{\text{audio-only tail}},
\end{equation*}
where $T$ is a BPE text token, $A$ is a 6.25~Hz audio position, and
$\langle\textsc{eot}\rangle$ marks the end of the text stream. The 1:1
interleaving lets an upstream conversational LLM drive synthesis at
its own text-emission rate: each text token it emits is consumed at
the next audio step, so speech can start within a single text token of
generation and continue token by token from there. The backbone then
serves directly as the audio output of a streaming dialogue LLM,
without buffering a full utterance before speaking. The
$\langle\textsc{eot}\rangle$ marker covers the common case where the
dialogue model finishes its text before the corresponding speech has
been fully rendered: audio continues from the already-received text
context until the stop head fires.

This mode is intended for real-time dialogue systems, where text and
audio are produced incrementally and synthesis should begin before a
full utterance is available. Plain mode remains the best choice when
the full text has already been prepared before synthesis.

\subsection{Semantic encoder}
\label{sec:model:semantic}

The semantic encoder is the same module that supervises Stage~2 of the
AudioVAE training, transplanted into the backbone with its pretrained
weights. It converts each newly generated 25~Hz VAE-latent patch into
a single 6.25~Hz embedding for the LLM, stripping out the high-variance
acoustic detail along the way.

Architecturally, a strided causal-convolution projector first halves
the input frame rate, followed by a 24-layer causal Transformer with
hidden dim $1024$ and FFN dim $4096$; the encoder output is then
grouped in pairs along time and linearly projected to the LLM embedding
dimension, yielding an end-to-end $4\times$ downsampling from 25~Hz
latent frames to 6.25~Hz LLM tokens. Strict causality at every step
lets the same module be unrolled one patch at a time during streaming
inference.

As the encoder is already pretrained to expose semantically
structured features, it maps a latent patch onto a representation
aligned with the LLM's text semantic space. The LLM can then ignore
acoustic detail during the AR rollout and condition on a compact
summary of the history.

\subsection{Autoregressive flow-matching head}
\label{sec:model:arfm}

\subsubsection{Per-step latent generation}

Following DiTAR and ARDiT, we use
a Diffusion Transformer (DiT)~\citep{dit} as the velocity-field
predictor, instantiated with 18 layers, hidden dim $1024$, and FFN dim
$4096$, with RoPE on every layer, RMSNorm with QK-norm, and
adaLN-zero modulation driven by the diffusion timestep and a
speaker-embedding side input. The training target is the rectified-flow
vector field of \citet{rectifiedflow}, regressed against the
straight-line velocity between Gaussian noise and the clean latent
patch. At inference we solve the ODE with a small number of Euler
steps.

Each per-step VAE patch is a block of 4 latent frames at 25~Hz
(matching the $4{:}1$ frame-rate ratio between the VAE latent and the
6.25~Hz LLM stream); we write $P_n = (\ell_n^1, \ldots, \ell_n^4)$ for
the clean patch and $Z_n = (z_n^1, \ldots, z_n^4)$ for its noisy
counterpart. At every audio position, the head consumes a
\emph{flow-matching context} assembled from three streams projected
into a common hidden space:
\begin{itemize}
    \item the LLM hidden state $H_n$ at that position (one token),
    \item the clean patches $P_{<n}$ of all earlier audio positions
    (four tokens each), and
    \item the noisy patch $Z_n$ under generation (four tokens).
\end{itemize}
These are interleaved into a single sequence
\begin{equation}
    [\,H_0,\,P_0,\,H_1,\,P_1,\,\ldots,\,H_{n-1},\,P_{n-1},\,H_n,\,Z_n\,],
    \label{eq:arfm-seq}
\end{equation}
where, with the convention above, each $H_n$ contributes one token and
each $P_n$ / $Z_n$ contributes a four-token block. The top-right inset
of \Cref{fig:overview} illustrates this layout together with the
substitution step that replaces $Z_n$ with the integrated $P_n$ in the
next-step history. A speaker x-vector
extracted by a frozen CAM++ encoder~\citep{campp} is added as a global
adaLN-zero condition, and the LLM hidden stream and the speaker stream
are each dropped independently with probability $p_{\mathrm{drop}}=0.5$
during training to enable classifier-free guidance~\citep{cfg} over
text content and timbre at inference. CFG is then applied jointly across
the two conditions: the conditional branch keeps both streams, the
unconditional branch drops both, and the velocity actually fed to the
ODE solver is the standard linear extrapolation of the two. At
inference, the AR loop runs as follows: starting from
$[H_0,\,Z_0]$ the integrator returns $P_0$; we substitute $P_0$ for
$Z_0$, append the next pair $[H_1,\,Z_1]$, integrate, and repeat. Each
newly produced $P_n$ is also fed back through the semantic encoder to
update the LLM's audio-semantic input for the next forward step.

\subsubsection{Block-causal training attention}

We train the AR-FM head in parallel across all $N$ patches of an
utterance with a single forward pass, while reproducing exactly the
per-step autoregressive context each patch would see at inference. The
mechanism is a block-causal attention mask over a concatenated
cause/generation sequence whose two halves share positional indices.

The training sequence is built from two halves of equal length. The
\emph{cause} part
\begin{equation*}
    C = [\,H_0, P_0, H_1, P_1, \ldots, H_{N}, P_{N}\,]
\end{equation*}
holds the clean LLM-conditioned history, and the \emph{generation}
part
\begin{equation*}
    Z = [\,H_0, Z_0, H_1, Z_1, \ldots, H_{N}, Z_{N}\,]
\end{equation*}
holds the noisy patches under denoising, each at an independently
sampled flow-matching time. Position indices are reset between the two halves:
tokens within $C$ are numbered $0, 1, \ldots, |C|-1$, and $Z$ restarts
from $0$ with the same per-block layout, so each $(H_n, Z_n)$ in $Z$
inherits the position range of the corresponding $(H_n, P_n)$ in $C$.
The RoPE phases between $Z_n$ and $P_{<n}$ therefore match those of a
per-step inference forward.

\begin{figure}[t]
    \centering
    \includegraphics[width=\linewidth]{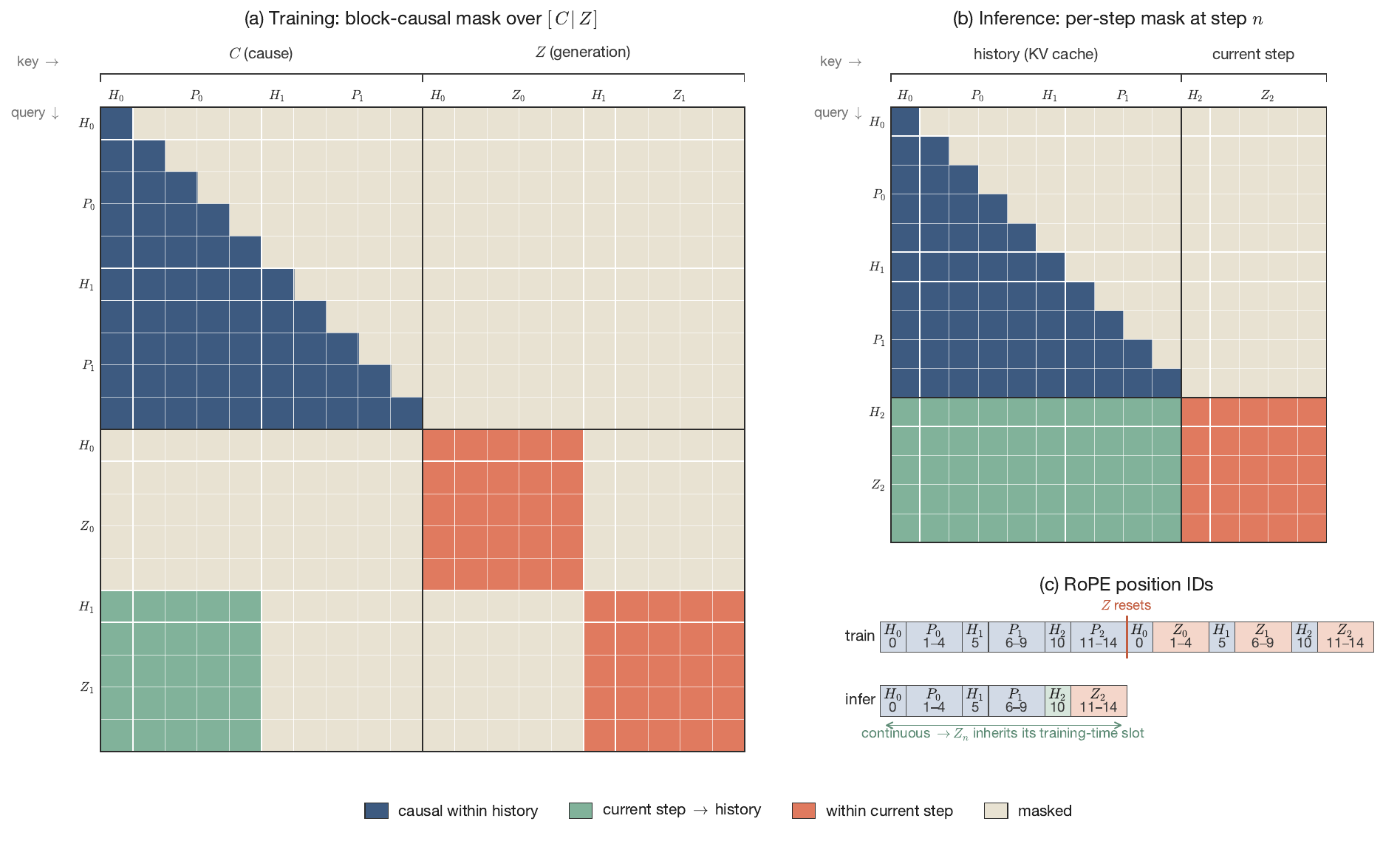}
    \caption{Attention masks and RoPE position IDs of the AR
    flow-matching head ($H$ blocks of size $1$, $P$/$Z$ blocks of size
    $L\!=\!4$). \textbf{(a)} Block-causal training mask over
    $[\,C\,|\,Z\,]$. \textbf{(b)} Per-step inference mask at step $n$,
    with the KV-cached history on the left and the newly appended
    $(H_n, Z_n)$ on the right. \textbf{(c)} Position IDs: the train row
    is the $C$ and $Z$ segments concatenated, with the red marker at
    the boundary where $Z$'s positions reset; the inference row places
    $Z_n$ at the same positions $Z_n$ occupied in training.}
    \label{fig:attn-mask}
\end{figure}

The block-causal mask partitions the $(C+Z)\times(C+Z)$ attention
matrix into four sub-blocks (\Cref{fig:attn-mask}): $C\!\to\!C$ is
standard causal, mirroring the LLM-side ordering of the prefix;
$C\!\to\!Z$ is fully masked, so the cause stream's hidden states are
independent of the noisy targets and equal to their inference-time
values; $Z\!\to\!C$ is prefix-causal, letting each generation block
$Z_n$ attend to exactly $H_{\le n}$ and $P_{<n}$, the autoregressive
context the patch sees at inference; and $Z\!\to\!Z$ is block-diagonal,
so different patches denoise independently under their own sampled
times. Combined with the shared positional indices, this layout makes
the parallel training forward numerically identical to a per-step
inference rollout while training all $N$ heads in one pass.

Because the AR-FM prefix already carries every LLM hidden state, the
AR-FM head is on its own a complete text-conditioned speech generator,
in the same sense as ARDiT: it could in principle
synthesize audio with no further acoustic feedback from the LLM. We
observe that this pushes the LLM toward encoding semantic rather than
acoustic information.

\subsection{Training objectives}
\label{sec:model:loss}

\subsubsection{AudioVAE objectives}
\label{sec:model:loss:vae}

The AudioVAE has its own two-stage objective, optimized ahead of the
backbone and frozen for every backbone stage below. Per training clip,
\begin{equation}
\mathcal{L}_{\mathrm{vae}}
=
\underbrace{
  \mathcal{L}_{\mathrm{mel}}
  + \mathcal{L}_{\mathrm{adv}}
  + \mathcal{L}_{\mathrm{fm}}
  + \beta_{\mathrm{kl}}\mathcal{L}_{\mathrm{kl}}
}_{\text{Stage 1}}
\;+\;
\underbrace{
  \lambda_{\mathrm{wavlm}}\mathcal{L}_{\mathrm{wavlm}}
  + \lambda_{\mathrm{sup}}\mathcal{L}_{\mathrm{sup}}
}_{\text{added in Stage 2}},
\label{eq:vae-loss}
\end{equation}
where the Stage 1 group is a BigVGAN-v2-style reconstruction stack
($\mathcal{L}_{\mathrm{mel}}$ multi-scale mel,
$\mathcal{L}_{\mathrm{adv}}$ multi-period plus sub-band CQT
adversarial, $\mathcal{L}_{\mathrm{fm}}$ their feature-matching
counterpart) regularized by a KL\,+\,flow prior on the
latent, and the Stage 2 group adds a frame-level cosine
alignment loss
$\mathcal{L}_{\mathrm{wavlm}}$ against a frozen WavLM
teacher together with a supervision loss
$\mathcal{L}_{\mathrm{sup}}$ summed over ASR, emotion, and speaker
heads routed through a downstream encoder--LLM block, whose encoder is
the module retained as the semantic frontend (\Cref{sec:model:semantic}).
All backbone losses below treat the AudioVAE as a fixed encoder/decoder
pair.

\subsubsection{Pretraining objectives}
\label{sec:model:loss:pretrain}

During backbone pretraining, two objectives drive the full model.

\paragraph{Flow-matching loss.}
A per-patch mean-squared error between the predicted velocity and the
analytic conditional vector field on the (sampled) VAE latent. Following the rectified-flow formulation, with
$P_n$ denoting the clean four-frame VAE-latent patch at audio
position $n$ and the per-step flow-matching context defined in
\Cref{sec:model:arfm}, we optimize
\begin{equation}
    \mathcal{L}_{\text{fm}}
    = \mathbb{E}_{n,\,t \sim \mathcal{U}(0,1),\,
                  \epsilon \sim \mathcal{N}(0, I),\, P_n}
      \left\|
        v_\theta\!\left(Z_n^t,\, t,\, H_{{\le}n},\, P_{<n},\, s\right)
        - \left(P_n - \epsilon\right)
      \right\|^2,
    \label{eq:loss-fm}
\end{equation}
where $\epsilon \sim \mathcal{N}(0, I)$ is sampled from a standard
Gaussian prior, $P_n$ is the ground-truth patch of four 25~Hz VAE
latent frames, $Z_n^t = (1 - t)\,\epsilon + t\,P_n$ is the linear
interpolation between the noise and the clean patch, $v_\theta$ is
the velocity field predicted by the DiT-based AR-FM head, $H_n$ is
the LLM hidden state at audio position $n$ that serves as the
per-step conditioning signal, $P_{<n}$ are the previously generated
clean patches forming the flow-matching prefix (cf.\
\Cref{eq:arfm-seq}), and $s$ is the global speaker x-vector. This is the \emph{only} loss that backpropagates through the LLM and the
semantic encoder: text positions carry no loss weight, audio-span
positions are routed to the flow-matching head, and the LLM is
therefore optimized end-to-end through the gradient flowing back from
the AR-FM head.

\paragraph{Stop loss.}
A dedicated EOS prediction head, a two-layer MLP attached on top of a
\emph{detached} copy of the LLM hidden state at every audio position,
is trained with a balanced binary cross-entropy that gives the (single)
positive position the same total mass as all earlier negatives. Let
$N$ be the number of audio positions in the utterance, with position
$N{-}1$ being the EOS target, and let
\begin{equation*}
    p_n = \sigma\!\left(\mathrm{MLP}\!\left(\operatorname{sg}(H_n)\right)\right)
\end{equation*}
denote the predicted stop probability at position $n$, where
$\operatorname{sg}(\cdot)$ is the stop-gradient operator and $\sigma$
is the sigmoid. The stop loss is then
\begin{equation}
    \mathcal{L}_{\text{eos}}
    = -\tfrac{1}{2}\,\log p_{N-1}
      \;-\; \frac{1}{2(N-1)} \sum_{n=0}^{N-2}
              \log\!\left(1 - p_n\right),
    \label{eq:loss-eos}
\end{equation}
so that the single positive at position $N{-}1$ contributes the same
total mass as the $N{-}1$ earlier negatives combined. Detaching the
hidden state keeps the stop predictor from interfering with the AR
rollout dynamics.

The total backbone pretraining loss is therefore
\begin{equation}
    \mathcal{L}_{\mathrm{pre}}
    = \mathcal{L}_{\mathrm{fm}} + \mathcal{L}_{\mathrm{eos}},
    \label{eq:pretrain-loss}
\end{equation}
with the two terms weighted equally. The AudioVAE and the speaker
encoder remain frozen throughout backbone training. Only the LLM, the
semantic encoder, the AR-FM head, the stop head, and the small
input/output projections that connect these modules receive gradients.

\subsubsection{Post-training objectives}
\label{sec:model:loss:posttrain}

After pretraining, we post-train only the DiT acoustic generator inside
the AR-FM head. The AudioVAE, speaker encoder, semantic encoder, and LLM
are frozen and provide fixed acoustic, speaker, and semantic context.
Let \(v_\theta\) denote the trainable velocity predictor in the
self-corrective stage, and let \(v_\phi\) denote the MeanFlow student in
the distillation stage. Superscript \((b)\) indexes a minibatch element,
where \(b=1,\ldots,B\), and \(B\) is the minibatch size. Unless otherwise
specified, squared norms are computed over the latent patch dimensions.

For notational simplicity, we use a compact noise-to-data notation in this
section, consistent with the flow-matching formulation in pretraining. For a
sampled VAE latent patch \(P_i\), we write
\[
    x_0 \equiv \epsilon \sim \mathcal{N}(0,I),\qquad
    x_1 \equiv P_i,\qquad
    x_t \equiv Z_i^t = (1-t)x_0 + t x_1,
\]
and denote all conditioning information by \(c\). Here, \(x_0\) is Gaussian
noise, \(x_1\) is the clean VAE latent patch, \(x_t\) is the interpolated
patch at flow time \(t\in[0,1]\), and \(c\) contains the fixed acoustic,
speaker, and semantic context provided by the frozen modules. The index
\(i\) denotes the sampled latent patch and is independent of the minibatch
index \(b\).

For any velocity predictor \(f\) and conditioning \(c\), classifier-free
guidance (CFG) is written as
\begin{equation}
    f^{\mathrm{cfg}}(x_t,t,c;\gamma)
    =
    f(x_t,t,c)
    +
    \gamma\left(f(x_t,t,c)-f(x_t,t,\varnothing)\right),
    \label{eq:posttrain-cfg}
\end{equation}
where \(\varnothing\) denotes the dropped-conditioning branch and
\(\gamma\ge 0\) is the CFG scale. When \(f=v_\theta\), we write
\(v_\theta^{\mathrm{cfg}}\) for the corresponding CFG-guided predictor.
This convention is used by both post-training stages.

\paragraph{Self-corrective alignment.}
Following SOAR, the first stage uses a reward-free
self-correction objective for the pretrained DiT. We use
\(\omega(t)\ge 0\) to denote the time-dependent regression weight for
flow matching, which balances the contribution of training samples at
different noise levels. The same weighting function is used for both
on-trajectory and auxiliary off-trajectory losses.

For each minibatch element \(b\), we sample a flow time
\(\tau^{(b)}\sim p_\tau\), where \(p_\tau\) is the training-time
distribution over \([0,1]\). For any time \(s\in[0,1]\), we write
\[
    x_s^{(b)} = (1-s)x_0^{(b)} + s x_1^{(b)} .
\]
We also introduce \(h_{\mathrm{soar}}>0\) as the self-correction rollout
step size in normalized flow time. Since our time convention moves from
noise at \(t=0\) to data at \(t=1\), a rollout from time \(\tau^{(b)}\)
advances to
\[
    \tau_+^{(b)}=\min(\tau^{(b)}+h_{\mathrm{soar}},1).
\]

The on-trajectory term is the usual flow-matching regression:
\begin{equation}
\ell_{\mathrm{on}}^{(b)}
=
\omega\!\left(\tau^{(b)}\right)
\left\|
    v_\theta\!\left(x_{\tau^{(b)}}^{(b)},\tau^{(b)},c^{(b)}\right)
    - \left(x_1^{(b)}-x_0^{(b)}\right)
\right\|_2^2 .
\label{eq:sca-on-loss}
\end{equation}

To expose the model to its own inference-time errors, we perform a
single detached Euler rollout from \(x_{\tau^{(b)}}^{(b)}\) using the
current CFG-guided predictor:
\begin{equation}
\begin{aligned}
    \hat{x}_{\tau_+}^{(b)}
    &=
    \operatorname{sg}\!\left[
        x_{\tau^{(b)}}^{(b)}
        +
        \left(\tau_+^{(b)}-\tau^{(b)}\right)
        v_\theta^{\mathrm{cfg}}
        \!\left(
            x_{\tau^{(b)}}^{(b)},
            \tau^{(b)},
            c^{(b)};
            \gamma_{\mathrm{soar}}
        \right)
    \right],
\end{aligned}
\label{eq:sca-rollout}
\end{equation}
where \(\gamma_{\mathrm{soar}}\ge 0\) is the CFG scale used during
self-corrective rollout. The stop-gradient operator
\(\operatorname{sg}[\cdot]\) prevents gradients from backpropagating
through the rollout trajectory.

We then re-noise this off-trajectory state toward the original noise
endpoint. Let \(K_{\mathrm{aux}}\) be the number of auxiliary samples drawn
for each minibatch element. For \(k=1,\ldots,K_{\mathrm{aux}}\),
\begin{equation}
\begin{aligned}
    \alpha^{(b,k)} &\sim \mathcal{U}(0,1), \\
    \tau_{\mathrm{aux}}^{(b,k)}
    &= (1-\alpha^{(b,k)})\tau_+^{(b)}, \\
    x_{\mathrm{aux}}^{(b,k)}
    &= \operatorname{sg}\!\left[
        (1-\alpha^{(b,k)})\hat{x}_{\tau_+}^{(b)}
        + \alpha^{(b,k)}x_0^{(b)}
    \right].
\end{aligned}
\label{eq:sca-aux-sampling}
\end{equation}
The auxiliary target is the endpoint-consistent velocity that transports
\(x_{\mathrm{aux}}^{(b,k)}\) to the clean endpoint \(x_1^{(b)}\) at
\(t=1\):
\begin{equation}
    u_{\mathrm{aux}}^{(b,k)}
    =
    \operatorname{sg}\!\left(
        \frac{x_1^{(b)}-x_{\mathrm{aux}}^{(b,k)}}
        {1-\tau_{\mathrm{aux}}^{(b,k)}+\varepsilon_{\mathrm{aux}}}
    \right),
\label{eq:sca-aux-target}
\end{equation}
where \(\varepsilon_{\mathrm{aux}}>0\) is a small numerical constant used
to avoid division by zero near the data endpoint.

The corresponding auxiliary loss is
\begin{equation}
\ell_{\mathrm{aux}}^{(b,k)}
=
\omega\!\left(\tau_{\mathrm{aux}}^{(b,k)}\right)
\left\|
    v_\theta\!\left(
        x_{\mathrm{aux}}^{(b,k)},
        \tau_{\mathrm{aux}}^{(b,k)},
        c^{(b)}
    \right)
    - u_{\mathrm{aux}}^{(b,k)}
\right\|_2^2 .
\label{eq:sca-aux-loss}
\end{equation}

Let
\[
    \mathcal{A}\subseteq
    \{1,\ldots,B\}\times\{1,\ldots,K_{\mathrm{aux}}\}
\]
denote the set of retained auxiliary samples, and let
\(M_{\mathcal A}=|\mathcal A|\). If no auxiliary samples are filtered,
then \(M_{\mathcal A}=B K_{\mathrm{aux}}\). With auxiliary-loss weight
\(\lambda_{\mathrm{aux}}\ge 0\), the self-corrective alignment loss is the
count-normalized combination
\begin{equation}
    \mathcal{L}_{\mathrm{soar}}
    =
    \frac{
        \sum_{b=1}^{B}\ell_{\mathrm{on}}^{(b)}
        +
        \lambda_{\mathrm{aux}}
        \sum_{(b,k)\in\mathcal{A}}\ell_{\mathrm{aux}}^{(b,k)}
    }{
        B+\lambda_{\mathrm{aux}}M_{\mathcal A}
    } .
\label{eq:sca-loss}
\end{equation}
This is the flow-matching-native post-training stage detailed in
\Cref{sec:exp:train:sca}; our self-corrective alignment applies
reward-free self-correction directly to the acoustic DiT rather than to
the full TTS stack.

\paragraph{CFG-aware MeanFlow distillation.}
The second stage freezes the self-corrected DiT as a teacher. We denote
this frozen teacher by \(v_{\theta_{\mathrm T}}\), and train a student
DiT \(v_\phi\) to predict the mean velocity over a time interval
\([t_a,t_b]\). The interval endpoints are sampled from a distribution
\(p_{\mathrm{mf}}\) over pairs satisfying
\[
    0 \le t_a < t_b \le 1,
    \qquad
    \Delta t = t_b-t_a .
\]
For a training sample \((x_0,x_1,c)\), the student input at the start of
the interval is
\[
    x_{t_a}=(1-t_a)x_0+t_a x_1 .
\]

Starting from
\[
    x_{t_a}^{\mathrm{T},\mathrm{cfg}} = x_{t_a},
\]
we generate a teacher trajectory over \([t_a,t_b]\) using the frozen
teacher with CFG scale \(\gamma_{\mathrm{mf}}\):
\[
    \frac{d x_t^{\mathrm{T},\mathrm{cfg}}}{dt}
    =
    v_{\theta_{\mathrm T}}^{\mathrm{cfg}}
    \!\left(x_t^{\mathrm{T},\mathrm{cfg}},t,c;\gamma_{\mathrm{mf}}\right),
    \qquad t\in[t_a,t_b].
\]
The teacher mean-velocity target is then approximated by the finite
difference
\begin{equation}
    \bar{v}^{\,\mathrm{T},\mathrm{cfg}}_{t_a\rightarrow t_b}
    \approx
    \frac{
        x^{\mathrm{T},\mathrm{cfg}}_{t_b}
        -
        x^{\mathrm{T},\mathrm{cfg}}_{t_a}
    }{
        t_b-t_a
    },
\label{eq:meanflow-target}
\end{equation}
where the approximation sign reflects the numerical integration used to
obtain \(x^{\mathrm{T},\mathrm{cfg}}_{t_b}\).

The student uses the same DiT backbone as the teacher, adds a duration
embedder for \(\Delta t\), and predicts the interval mean velocity with a
single conditional forward pass:
\[
    v_\phi(x_{t_a},t_a,\Delta t,c).
\]
It is trained with
\begin{equation}
\begin{aligned}
    \mathcal{L}_{\mathrm{mv}}
    &=
    \mathbb{E}_{(x_0,x_1,c),\, (t_a,t_b)}
    \!\left[
        w_{\mathrm{mv}}\,\ell_{\mathrm{mv}}
    \right],\\
    \ell_{\mathrm{mv}}
    &=
    \left\|
        v_\phi(x_{t_a},t_a,\Delta t,c)
        -
        \bar{v}^{\,\mathrm{T},\mathrm{cfg}}_{t_a\rightarrow t_b}
    \right\|_2^2 .
\end{aligned}
\label{eq:meanflow-loss}
\end{equation}
We use the adaptive per-sample weight
\begin{equation}
    w_{\mathrm{mv}}
    =
    \left(
        \operatorname{sg}(\ell_{\mathrm{mv}})
        +
        \varepsilon_{\mathrm{mv}}
    \right)^{-1/2},
\label{eq:meanflow-weight}
\end{equation}
where \(\varepsilon_{\mathrm{mv}}>0\) is a small numerical constant. The
stop-gradient in \(w_{\mathrm{mv}}\) prevents gradients from flowing
through the adaptive weighting term.

Because CFG is fused into the teacher target, the student directly
matches the teacher's CFG-guided mean-velocity prediction with a single
conditional forward pass at inference, avoiding the separate conditional
and unconditional evaluations required by standard CFG while preserving
the corrected teacher behavior.

\section{Experiments}
\label{sec:exp}

This section summarizes the \dotstts{} training recipe and evaluates four
axes: foundational quality, multilingual coverage, cross-lingual voice
cloning, and expressiveness. We use Seed-TTS-Eval,
the MiniMax-Speech multilingual test set,
CV3-Eval, and EmergentTTS-Eval.

\subsection{Data}
\label{sec:exp:data}

In total, the backbone is trained on 1.5M hours of audio
across speech, captioned speech, and a small fraction of general audio,
drawn from three sources: an in-house Chinese/English speech pool, a
curated mixture of open-source TTS and ASR corpora, and a small
caption-paired set, described in turn below.

\paragraph{In-house data.}
The bulk of our training audio comes from an internal Chinese and
English speech corpus. A unified preprocessing stack applies vocal
enhancement, source separation, speaker-aware diarization, and
language-routed ASR: Whisper-Large-v3~\citep{whisper} for English and
most other languages, and Paraformer~\citep{paraformer} for Mandarin
Chinese. We then filter clips with cross-ASR consistency,
effective-bandwidth estimation, UTMOS, and intra-clip x-vector
variance. After filtering, the set contains approximately 1.2M hours of
cleaned, transcribed, and speaker-organized speech for backbone
training.

\paragraph{Open-source corpora.}
Alongside the in-house pool, we incorporate a curated mixture of
open-source TTS and ASR corpora that meet our quality bar after
re-running the same scoring stack. The mixture includes Emilia,
LibriTTS-R, HiFi-TTS, HiFi-TTS-2, WenetSpeech4TTS, AISHELL-3,
Magicdata, MLS, MSR-86K, IndicVoices-R, EuroSpeech, WaxalNLP-TTS, and
FLEURS, totalling roughly 300K hours and contributing the bulk of our
non-CJK language coverage.

\paragraph{Caption-style data.}
To bootstrap natural-language style control and general-audio
generation, we also curate a small caption-paired set. It combines a
sample of AutoACD~\citep{autoacd} with natural-language captions and
an in-house subset of open-source corpora augmented with Gemini-generated
descriptions of speaker traits, emotion, delivery, and acoustic
environment. This caption-paired set totals approximately 7K hours.

\subsection{Training}
\label{sec:exp:train}

The full training pipeline proceeds in three stages: pretraining on
the mixture of \Cref{sec:exp:data}, flow-matching-native post-training,
and MeanFlow distillation for low-NFE inference. Detailed recipes for
each stage are deferred to the following subsections.

\subsubsection{AudioVAE}
\label{sec:exp:train:audiovae}

The AudioVAE is trained on 48~kHz audio drawn from the in-house and
open-source pools of \Cref{sec:exp:data}, augmented with a small
general-audio share so the latent covers non-speech sounds as well.
Optimization uses AdamW with $\beta=(0.8,0.99)$, $\epsilon=10^{-6}$,
and an exponential learning-rate decay from $10^{-4}$ to $10^{-6}$.
Stage 1 runs for 500K steps on 9.6-second cropped segments. Stage 2 runs for a further 200K steps and
matches the latent against the 23rd-layer hidden representation of
WavLM as the frame-level teacher. The resulting AudioVAE is frozen
for the rest of this section.

\subsubsection{Pretraining}
\label{sec:exp:train:pretrain}

Pretraining proceeds in three stages (modality alignment, general
training, and annealing) that progressively widen the trainable
parameter set, the training pool, and the data difficulty.
Throughout all three stages we use the AdamW
optimizer under a single Warmup-Stable-Decay (WSD) schedule: each stage
opens with its own linear warmup from $2\times10^{-6}$ over the first
1\% of that stage's steps to the shared peak learning rate of
$2\times10^{-4}$, and the decay phase is deferred to the final
annealing stage, where the learning rate is linearly decayed from
$2\times10^{-4}$ to $3\times10^{-5}$.

\paragraph{Stage 1: modality alignment.}
The first stage opens a usable channel between the audio representation
and the LLM's semantic space. We freeze the LLM backbone and update
only the semantic encoder and the AR-FM head on an internal GPU
cluster with a global batch size of approximately 0.5~hours of audio,
training for 100K optimization steps. We restrict the data to Emilia in this stage:
pilot runs on the full pretraining mixture were severely unstable, and
we attribute this to the cost of routing gradients through encoders
that have not yet aligned with a frozen LLM. We also observe that, despite the LLM being held fixed, the backbone
already produces intelligible speech end-to-end.
Pronunciation is frequently incorrect and the resulting audio scores
around 42\% WER on Seed-TTS-Eval, but the alignment channel is
unambiguously open before any LLM parameter is touched.

\paragraph{Stage 2: general training.}
Once the modality channel is established, we unfreeze every module and
train end-to-end on the full data mixture of \Cref{sec:exp:data} without
any explicit reweighting between sources. This stage runs with a global
batch size of approximately 8~hours of audio for 700K steps,
corresponding to four epochs over the corpus. Its role is to build the
text-token-to-audio mapping at scale and to absorb the bulk of the 
cross-lingual, multi-domain coverage of the training pool.

\paragraph{Stage 3: annealing.}
The final stage anneals on a higher-quality subset obtained by
re-filtering the general training pool with a tighter threshold on the
in-house quality score described in \Cref{sec:exp:data}. This stage
runs for 100K steps, approximately one epoch
over the filtered subset, during which the WSD schedule enters its
decay phase and the learning rate is linearly annealed from the peak
$2\times10^{-4}$ down to $3\times10^{-5}$. All evaluation numbers
reported under the \textit{\dotstts{} (Pretrain)} rows in the remainder
of this section are produced with the checkpoint at the end of this
stage.

\subsubsection{Self-corrective alignment}
\label{sec:exp:train:sca}
\label{sec:exp:train:posttrain}

The self-corrective alignment stage updates only the DiT acoustic generator in the
AR-FM head. The LLM, semantic encoder, AudioVAE, and speaker encoder are
kept frozen. We use the self-corrective alignment objective in
\Cref{eq:sca-loss} on the same multilingual / multi-dialect training
pool as pretraining, filtered to high-quality utterances. Training runs
with a global batch containing 4 hours of audio
for 50K optimization steps with a 5K-step linear warmup to a peak
learning rate of $3\times10^{-5}$, followed by cosine decay to
$2\times10^{-6}$. The auxiliary correction weight is
$\lambda_{\mathrm{aux}}=1.0$. The one-step rollout uses CFG scale
$\gamma_{\mathrm{soar}}=1.2$, and each main sample generates $N=6$
auxiliary correction states. Integration times follow a logit-normal
sampling distribution under the noise-to-data convention of
\Cref{sec:model:loss:posttrain}.

This stage is reward-free: it does not use a reward model, human
preference model, or external acoustic teacher. Instead, the current DiT
constructs its own detached off-trajectory states with
\Cref{eq:sca-rollout,eq:sca-aux-sampling}, and the supervised correction
target in \Cref{eq:sca-aux-target} teaches the same DiT to steer those
states back toward the clean latent endpoint. In practice this directly
addresses the multi-step ODE mismatch between pretraining and inference,
where small velocity errors accumulate across AR patches. Throughout this paper, \textit{\dotstts{} (SOAR)} denotes the checkpoint obtained
after this self-corrective post-training stage.

\subsubsection{MeanFlow distillation}
\label{sec:exp:train:meanflow}

After self-corrective alignment, we freeze the corrected DiT as the
teacher and initialize a student DiT from it. The student keeps the same
backbone and adds only the interval-duration embedder used by
\Cref{eq:meanflow-loss}; all compatible parameters are copied from the
teacher, while the duration embedder is initialized separately. Teacher
trajectories are constructed with a 16-step Euler solver, and CFG is
applied when forming the teacher target so that the student learns the
CFG-guided mean-velocity prediction directly.

The MeanFlow (MF) stage uses the same setup and 8-hour global audio
batch, and is trained for 50K steps with a 5K-step warmup and cosine
decay to $2\times10^{-6}$. We use a larger peak learning rate of
$1\times10^{-4}$ because the student learns an interval-conditioned
mean-velocity objective rather than the original instantaneous velocity.
Intervals are sampled from a log-normal distribution with mean $-0.4$
and standard deviation $1.0$ in the same noise-to-data time
parameterization. Following the MeanFlow recipe, anchor samples are mixed
with probability $0.5$ for optimization stability. At inference, the
student updates intervals as
$x_{t_b}=x_{t_a}+\Delta t\,v_\phi(x_{t_a},t_a,\Delta t,c)$, with
$\Delta t=t_b-t_a$, and needs only one
conditional forward pass per step because CFG has already been fused into
the distillation target. The resulting MeanFlow-distilled checkpoint is denoted as \textit{\dotstts{} (MF)} and is used for all evaluations labeled \textit{\dotstts{} (MF)} in the
remainder of this section.

\subsection{Evaluation}
\label{sec:exp:eval}

\subsubsection{Setup}
\label{sec:exp:eval:setup}

\paragraph{Metrics.}
Throughout this section, content fidelity is measured by ASR-based
word/character error rate (WER for English and other non-Chinese
languages, CER reported as WER for Chinese for table compatibility),
and speaker similarity (SIM) is the cosine similarity between
WavLM-SV embeddings of the generated speech and the
corresponding reference prompt. We follow the de facto ASR convention
used by recent reports: Whisper-Large-v3 for English and multilingual
benchmarks, Paraformer for Mandarin Chinese, and the per-benchmark
default ASR for the remaining languages on the MiniMax multilingual
set. All numbers reported under the \textit{\dotstts{} (Pretrain)} row are
produced with the pretrained checkpoint from
\Cref{sec:exp:train:pretrain}; the MeanFlow-distilled student is
reported in the inference-efficiency section
(\Cref{sec:efficiency}).

\paragraph{Inference configuration.}
\dotstts{}-Pretrain and \dotstts{}-SOAR use 10 Euler solver steps
with CFG, i.e., each step performs one conditional and one
unconditional DiT forward pass for an effective NFE of 20. In contrast,
\dotstts{}-MF is evaluated at NFE~$\in\{2, 3, 4\}$ with no CFG:
the guidance has already been fused into the MeanFlow distillation target,
so each step uses a single conditional forward pass. The CFG scale is
$\gamma = 1.2$ for CFG in \dotstts{}-Pretrain and
\dotstts{}-SOAR, and for the teacher target used to train \dotstts{}-MF
(\Cref{eq:posttrain-cfg}). All runs use float32 and have
\emph{torch.compile} disabled.

\paragraph{Baselines.}
We compare against a broad set of recent zero-shot TTS systems,
spanning both open-source releases and commercial products:
CosyVoice~2~\citep{cosyvoice2}, CosyVoice~3~\citep{cosyvoice3},
Qwen3-TTS~\citep{qwen3tts}, IndexTTS~2~\citep{indextts2},
FireRedTTS-2~\citep{firered2}, Seed-TTS~\citep{seedtts},
MiniMax-Speech~\citep{minimaxspeech},
Fish-Audio~S2~\citep{fishaudios2}, F5-TTS~\citep{f5tts},
MegaTTS3~\citep{megatts3}, DiTAR~\citep{ditar},
VibeVoice~\citep{vibevoice}, and VoxCPM~2~\citep{voxcpm2}. All baseline
numbers reported in this section are taken either from the original
papers or from the official default inference configuration of the
corresponding open-source release; we do not re-tune any baseline.

\begin{table}[t]
\centering
\small
\setlength{\tabcolsep}{5pt}
\caption{Speech reconstruction performance on \emph{LibriSpeech
test-other}. FPS is the
temporal frame rate of the underlying representation.
\textbf{Bold} marks the best non-oracle entry per column. ``--''
indicates a metric not reported by the corresponding source.}
\label{tab:audiovae}
\begin{tabular}{l c c cc c c c c}
\toprule
\multirow{2}{*}{\textbf{Model}}
  & \multirow{2}{*}{\textbf{Sample Rate}}
  & \multirow{2}{*}{\textbf{FPS}}
  & \multicolumn{2}{c}{\textbf{PESQ$\uparrow$}}
  & \multirow{2}{*}{\textbf{STOI$\uparrow$}}
  & \multirow{2}{*}{\textbf{UTMOS$\uparrow$}}
  & \multirow{2}{*}{\textbf{SIM$\uparrow$}}
  & \multirow{2}{*}{\textbf{WER(\%)$\downarrow$}} \\
\cmidrule(lr){4-5}
 & & & NB & WB & & & & \\
\midrule
Ground Truth     & --    & --    & 4.55 & 4.64 & 1.000 & 3.50  & 1.000 & 4.59 \\
\midrule
\multicolumn{9}{l}{\textit{Discrete tokens}} \\
XY-Tokenizer     & 16~kHz & 100   & 2.80 & 2.23 & 0.89  & 3.46  & 0.82  & 6.19 \\
WavTokenizer     & 16~kHz & 75    & 2.40 & 1.96 & 0.87  & 3.22  & 0.68  & 13.35 \\
X-codec2         & 16~kHz & 50    & 2.83 & 2.26 & 0.90  & 3.64  & 0.81  & 6.85 \\
SAC              & 16~kHz & 62.5  & 2.92 & 2.39 & 0.90  & \textbf{3.84} & 0.85 & 5.77 \\
\midrule
\multicolumn{9}{l}{\textit{Continuous representations}} \\
SemanticVAE      & 16~kHz & 40    & 3.99 & 3.80 & 0.969 & 3.76  & 0.963 & 4.15 \\
MingTok-Audio    & 16~kHz & 50    & \textbf{4.23} & \textbf{4.12} & \textbf{0.981} & 3.75 & 0.950 & 4.27 \\
VibeVoice        & 24~kHz & \textbf{7.5}   & 2.85 & --  & 0.823 & 3.72  & --   & --  \\
\midrule
\textbf{\dotstts{} VAE} & \textbf{48~kHz} & 25 & 4.09 & 3.95 & 0.973 & 3.75 & \textbf{0.969} & \textbf{4.14} \\
\bottomrule
\end{tabular}
\end{table}

\subsubsection{Reconstruction Quality of the AudioVAE}
\label{sec:exp:eval:audiovae}

Before turning to the end-to-end TTS benchmarks, we first measure the
AudioVAE's reconstruction performance. \Cref{tab:audiovae} reports waveform-reconstruction metrics on \emph{LibriSpeech test-other} 
against representative discrete neural codecs
(XY-Tokenizer~\citep{xytokenizer}, WavTokenizer~\citep{wavtokenizer},
X-codec2~\citep{xcodec2}, SAC~\citep{sac}) and continuous
representations (SemanticVAE~\citep{semanticvae},
MingTok-Audio~\citep{mingtok}, and the continuous tokenizer of
VibeVoice~\citep{vibevoice}). All baseline numbers are taken from the
corresponding original publications.

The discrete tokenizers in the upper block trail the continuous
representations on PESQ, STOI, SIM, and WER. Within the continuous
group, the \dotstts{} VAE has the highest input bandwidth and
second-lowest frame rate in the table, with PESQ, STOI, UTMOS,
SIM, and WER all in the top band of the group. The only other
low-frame-rate continuous entry, VibeVoice at 7.5~Hz, drops to
PESQ-NB 2.85 and STOI 0.823.

At WER 4.14 and SIM 0.969, the latent itself adds almost nothing
to end-to-end error, so reconstruction is not a downstream
bottleneck. Stage 2 (\Cref{sec:model:vae}) further adds
WavLM-aligned and multitask supervision that makes the latent
learnable as an LLM generation target. Both contribute to the
downstream TTS results in the remainder of this section.

\begin{table}[t]
\centering
\small
\setlength{\tabcolsep}{5pt}
\caption{Zero-shot results on Seed-TTS-Eval.
\textbf{Bold} marks the best per column.}
\label{tab:seedtts}
\begin{tabular}{l c cc cc cc cc}
\toprule
\multirow{2}{*}{\textbf{Model}} & \multirow{2}{*}{\textbf{Params}}
  & \multicolumn{2}{c}{\textbf{test-en}}
  & \multicolumn{2}{c}{\textbf{test-zh}}
  & \multicolumn{2}{c}{\textbf{test-zh-hard}}
  & \multicolumn{2}{c}{\textbf{Average}} \\
\cmidrule(lr){3-4} \cmidrule(lr){5-6} \cmidrule(lr){7-8} \cmidrule(lr){9-10}
 & & WER(\%)$\downarrow$ & SIM$\uparrow$
   & WER$\downarrow$ & SIM$\uparrow$
   & WER$\downarrow$ & SIM$\uparrow$
   & WER$\downarrow$ & SIM$\uparrow$ \\
\midrule
CosyVoice~3                  & 1.5B & 2.22 & 72.0 & 1.12 & 78.1 & \textbf{5.83} & 75.8 & 3.06          & 75.3 \\
F5-TTS                       & 0.3B & 2.00 & 67.0 & 1.53 & 76.0 & 8.67 & 71.3 & 4.10          & 71.4 \\
FireRedTTS~2                 & 1.5B & 1.95 & 66.5 & 1.14 & 73.6 & 8.98 & 70.3 & 4.02          & 70.1 \\
IndexTTS~2                   & 1.5B & 2.23 & 70.6 & 1.03 & 76.5 & 7.12 & 75.5 & 3.46          & 74.2 \\
MegaTTS~3                    & 0.5B & 2.79 & \textbf{77.1} & 1.52 & 79.0 & ---  & ---  & ---           & ---  \\
MiniMax-Speech               & ---  & 1.65 & 69.2 & \textbf{0.83} & 78.3 & ---  & ---  & ---           & ---  \\
Qwen3-TTS                    & 1.7B & \textbf{1.23} & 71.7 & 1.22 & 77.0 & 6.76 & 74.8 & 3.07          & 74.5 \\
Seed-TTS                     & ---  & 2.25 & 76.2 & 1.12 & 79.6 & 7.59 & 77.6 & 3.65          & 77.8 \\
DiTAR                        & 0.6B & 1.69 & 73.5 & 1.02 & 75.3 & ---  & ---  & ---           & ---  \\
VibeVoice                    & 1.5B & 3.04 & 68.9 & 1.16 & 74.4 & ---  & ---  & ---           & ---  \\
VoxCPM~2                     & 2B   & 1.84 & 75.3 & 0.97 & 79.5 & 8.13 & 75.3 & 3.65          & 76.7 \\
\midrule
\textbf{\dotstts{} (Pretrain)}    & 2B & 1.34 & 76.8          & 0.96 & 80.5          & 6.46 & 79.2          & \textbf{2.92} & 78.8 \\
\textbf{\dotstts{} (SOAR)}         & 2B & 1.30 & \textbf{77.1} & 0.94 & \textbf{81.0} & 6.60 & \textbf{79.5} & 2.95          & \textbf{79.2} \\
\quad -- MF, NFE=4            & 2B & 1.29 & 76.2          & 0.94 & 80.0          & 6.60 & 78.5          & 2.94          & 78.2 \\
\quad -- MF, NFE=3            & 2B & 1.41 & 75.9          & 1.02 & 79.9          & 7.19 & 78.6          & 3.21          & 78.1 \\
\quad -- MF, NFE=2            & 2B & 1.51 & 75.2          & 1.04 & 79.1          & 7.74 & 76.7          & 3.43          & 77.0 \\
\bottomrule
\end{tabular}
\end{table}

\subsubsection{Seed-TTS-Eval}
\label{sec:exp:eval:seedtts}

Seed-TTS-Eval is a widely used zero-shot
voice-cloning benchmark and the primary point of comparison in
this report. We evaluate on all three subsets (test-en, test-zh,
test-zh-hard) under the standard zero-shot protocol: the model
is conditioned on an approximately 3-second reference prompt
unseen during training, generates the target transcript, and is
scored with the benchmark's reference ASR and WavLM-SV
similarity.

\Cref{tab:seedtts} reports the results. \dotstts{} takes the best
average on both metrics. On SIM, the SOAR row leads at 79.2,
1.4 above the next-best baseline Seed-TTS (77.8) and 2.5 above
VoxCPM~2 (76.7). On WER, the Pretrain (2.92\%), SOAR (2.95\%), and
MF NFE\,$=4$ (2.94\%) rows all sit below every reported baseline,
with CosyVoice~3 next at 3.06\%.

The post-training stage of \Cref{sec:exp:train:sca} corrects the
multi-step ODE mismatch between pretraining and inference. The
Pretrain~$\to$~SOAR rows in the table reflect this correction
with broadly improved numbers. MeanFlow distillation at
NFE\,$=4$ keeps WER within 0.01 of SOAR on every column at a
cost of about one point of SIM. We also report MF NFE\,$=2$ and
NFE\,$=3$ in the table for comparison, with NFE\,$=4$ as the
best operating point of the three.

\subsubsection{MiniMax-multilingual Test Set}
\label{sec:exp:eval:minimax}

The MiniMax-Speech multilingual test set
covers 24 languages with 100 utterances per language and two MCV
reference speakers per language, and is widely used as a
multilingual zero-shot benchmark. We evaluate the full 24-language
set with the benchmark's per-language ASR.

\Cref{tab:minimax} reports per-language WER and SIM. \dotstts{}
(SOAR) leads the average SIM at 83.9, 1.6 above the next-best
baseline VoxCPM~2 (82.3), and a \dotstts{} variant takes the
per-language SIM lead outright on 19 of 24 languages and ties on
2 more.

The picture on WER is mixed. \dotstts{} is competitive on most
languages, but the average is pulled up by a few low-resource
outliers. Since SIM stays in the normal band on these languages, the
WER gap points to insufficient BPE token coverage
(\Cref{sec:model:llm}). Means to close the low-resource WER gap
are discussed in \Cref{sec:limitations}.

MeanFlow distillation (MF$_4$) matches SOAR at the average level
(6.8\%\,/\,83.5 vs 6.8\%\,/\,83.9), with occasional per-language wins
and about 0.4 SIM cost on the average row.

\begin{table}[t]
\centering
\small
\setlength{\tabcolsep}{4pt}
\caption{Per-language WER\,/\,SIM on the MiniMax-Speech multilingual
test set. \textbf{Bold} marks the best per
metric per row. The Average row is computed only over the languages
where each system reports a number.
$^*$Cantonese WER reflects an ASR-faithfulness floor common to all
systems in the table; the SIM column remains comparable.}
\label{tab:minimax}
\resizebox{\textwidth}{!}{%
\begin{tabular}{lccccccc}
\toprule
\textbf{Language} & \textbf{MiniMax} & \textbf{ElevenLabs} & \textbf{Fish-Audio S2} & \textbf{VoxCPM 2} & \textbf{\dotstts{} (Pre.)} & \textbf{\dotstts{} (SOAR)} & \textbf{\dotstts{} (MF$_4$)} \\
                  & WER(\%) / SIM        & WER / SIM           & WER / SIM              & WER / SIM        & WER / SIM                & WER / SIM                & WER / SIM \\
\midrule
Arabic       & \textbf{1.67} / 73.6 & \textbf{1.67} / 70.6 & 3.50 / 75.0          & 13.05 / \textbf{79.1}         & 37.91 / 77.5         & 36.19 / \textbf{79.1} & 39.65 / 77.6 \\
Cantonese$^*$ & 34.11 / 77.8        & 51.51 / 67.0         & 30.67 / 80.5         & 38.58 / 83.5                  & 37.91 / 84.7         & 42.32 / \textbf{85.0} & 37.82 / 84.0 \\
Chinese      & 2.25 / 78.0          & 16.03 / 67.7         & \textbf{0.73} / 81.6 & 1.14 / \textbf{82.5}          & 1.08 / 82.3          & 0.77 / \textbf{82.5}  & 1.01 / 81.8 \\
Czech        & 3.88 / 79.6          & \textbf{2.11} / 68.5 & 2.84 / 79.8          & 24.13 / 78.3                  & 5.05 / 83.8          & 4.25 / \textbf{84.2}  & 5.67 / 83.9 \\
Dutch        & 1.14 / 73.8          & \textbf{0.80} / 68.0 & 0.99 / 73.0          & 0.91 / 80.8                   & 1.20 / 81.4          & 1.39 / \textbf{82.2}  & 1.30 / 82.1 \\
English      & 2.16 / 75.6          & 2.34 / 61.3          & 1.62 / 79.7          & 2.29 / 85.4                   & 1.06 / 86.9          & \textbf{1.03} / \textbf{87.5} & 1.09 / 86.9 \\
Finnish      & 4.67 / 83.5          & 2.96 / 75.9          & 3.33 / 81.9          & \textbf{2.63} / \textbf{89.0} & 3.44 / 88.0          & 4.08 / 88.3           & 3.61 / 88.3 \\
French       & 4.10 / 62.8          & 5.22 / 53.5          & \textbf{3.05} / 69.8 & 4.53 / 73.5                   & 3.82 / 78.2          & 3.56 / \textbf{78.6}  & 3.26 / 78.5 \\
German       & 1.91 / 73.3          & 0.57 / 61.4          & \textbf{0.55} / 76.7 & 0.68 / 80.3                   & 1.03 / 79.5          & 1.70 / \textbf{80.6}  & 0.91 / 79.5 \\
Greek        & 2.02 / 82.6          & \textbf{0.99} / 73.3 & 5.74 / 79.5          & 2.84 / 86.0                   & 2.97 / \textbf{87.6} & 3.00 / \textbf{87.6}  & 3.19 / 87.3 \\
Hindi        & 6.96 / 81.8          & \textbf{5.83} / 73.0 & 14.64 / 82.1         & 19.70 / \textbf{85.6}         & 14.32 / 84.5         & 14.24 / 84.7          & 14.75 / 84.8 \\
Indonesian   & 1.24 / 72.9          & \textbf{1.06} / 66.0 & 1.46 / 76.3          & 1.08 / 80.0                   & 2.71 / 80.8          & 2.96 / 80.8           & 3.91 / \textbf{81.2} \\
Italian      & 1.54 / 69.9          & 1.74 / 57.9          & \textbf{1.27} / 74.7 & 1.56 / 78.0                   & 3.16 / 84.5          & 3.12 / \textbf{84.7}  & 2.16 / 84.3 \\
Japanese     & 3.52 / 77.6          & 10.65 / 73.8         & \textbf{2.76} / 79.6 & 4.63 / 82.8                   & 7.16 / 83.1          & 5.28 / \textbf{83.7}  & 5.17 / 83.1 \\
Korean       & 1.75 / 77.6          & 1.87 / 70.0          & \textbf{1.18} / 81.7 & 1.96 / 83.3                   & 5.30 / 84.3          & 5.66 / 83.6           & 3.93 / \textbf{84.9} \\
Polish       & 1.42 / 80.2          & \textbf{0.77} / 72.9 & 1.26 / 81.9          & 1.14 / \textbf{88.4}          & 2.72 / 87.3          & 3.59 / 87.8           & 3.42 / 87.5 \\
Portuguese   & 1.88 / 80.5          & 1.33 / 71.1          & \textbf{1.14} / 78.1 & 1.94 / 83.7                   & 1.64 / 83.1          & 2.00 / \textbf{84.3}  & 2.40 / 83.1 \\
Romanian     & 2.88 / 80.9          & \textbf{1.35} / 69.9 & 10.74 / 73.3         & 21.58 / 79.7                  & 3.36 / 86.2          & 3.87 / \textbf{87.1}  & 3.38 / 86.1 \\
Russian      & 4.28 / 76.1          & 3.88 / 67.6          & \textbf{2.40} / 79.0 & 3.63 / 81.1                   & 3.64 / 83.0          & 4.28 / \textbf{83.2}  & 4.42 / \textbf{83.2} \\
Spanish      & 1.03 / 76.2          & 1.08 / 61.5          & 0.91 / 77.6          & 1.44 / 83.1                   & 0.96 / 83.9          & 1.27 / \textbf{84.0}  & \textbf{0.80} / \textbf{84.0} \\
Thai         & \textbf{2.70} / 80.0 & 73.94 / 58.8         & 4.23 / 78.6          & 2.96 / 84.0                   & 7.45 / 83.8          & 7.86 / 83.9           & 8.03 / \textbf{84.2} \\
Turkish      & 1.52 / 77.9          & \textbf{0.70} / 59.6 & 0.87 / 83.5          & 0.82 / 87.1                   & 5.45 / \textbf{87.4} & 4.96 / 87.3           & 6.20 / 86.8 \\
Ukrainian    & 1.08 / 73.0          & \textbf{1.00} / 64.7 & 2.30 / 74.7          & 6.32 / 79.8                   & 1.61 / 80.5          & 1.27 / \textbf{81.2}  & 1.66 / 80.0 \\
Vietnamese   & \textbf{0.88} / 74.3 & 73.42 / 36.9         & 7.41 / 74.0          & 3.31 / 80.6                   & 3.85 / 80.7          & 3.89 / \textbf{81.6}  & 5.43 / 80.5 \\
\midrule
\textbf{Average} & \textbf{2.8} / 76.6 & 7.5 / 65.5         & 3.7 / 78.0           & 5.7 / 82.3                    & 6.6 / 83.5           & 6.8 / \textbf{83.9}   & 6.8 / 83.5 \\
\bottomrule
\end{tabular}%
}
\end{table}

\subsubsection{CV3-Eval}
\label{sec:exp:eval:cv3}

CV3-Eval, released with CosyVoice~3, complements the
previous two benchmarks with a hard-subset Chinese/English split and,
importantly, a \emph{cross-lingual} voice-cloning subset that uses a
reference speaker from one language to generate text in another.
Cross-lingual cloning is one of the hardest tests of timbre
disentanglement.

\begin{table}[t]
\centering
\small
\setlength{\tabcolsep}{5pt}
\caption{Results on CV3-Eval. W $=$ WER, S $=$ SIM.}
\label{tab:cv3}
\begin{tabular}{l cccc cc cc}
\toprule
\multirow{2}{*}{\textbf{Model}}
  & \multicolumn{4}{c}{\textbf{Monolingual} W(\%)$\downarrow$}
  & \multicolumn{2}{c}{\textbf{en$\to$zh}}
  & \multicolumn{2}{c}{\textbf{zh$\to$en}} \\
\cmidrule(lr){2-5} \cmidrule(lr){6-7} \cmidrule(lr){8-9}
 & zh & en & hard-zh & hard-en
 & W$\downarrow$ & S$\uparrow$
 & W$\downarrow$ & S$\uparrow$ \\
\midrule
CosyVoice~2                           & 4.08 & 6.32 & 12.58 & 11.96 & 13.50 & 63.3 & 6.47 & 64.3 \\
CosyVoice~3 (1.5B)                    & 3.91 & 4.99 & 9.77  & 10.55 & \textbf{8.01}  & 66.9 & \textbf{4.32} & 66.4 \\
Fish-Audio S2                         & \textbf{2.65} & \textbf{2.43} & 9.10  & 4.40           & ---   & ---  & ---  & ---  \\
VoxCPM~2                              & 3.65 & 5.00 & \textbf{8.55}  & 8.48  & ---   & ---  & ---  & ---  \\
\midrule
\textbf{\dotstts{} (Pretrain)}          & 3.51 & 5.24 & 9.69 & 5.99           & 10.88 & 74.6          & 4.97 & 71.9          \\
\textbf{\dotstts{} (SOAR)}               & 3.71 & 4.50 & 9.22 & 4.49           & 10.75 & \textbf{75.0} & 5.66 & \textbf{72.8} \\
\textbf{\dotstts{} (MF, NFE=4)}         & 3.95 & 4.05 & 9.10 & \textbf{4.37}  & 10.73 & 73.8          & 5.24 & 70.9          \\
\bottomrule
\end{tabular}
\end{table}

\Cref{tab:cv3} reports the CV3-Eval numbers. \dotstts{} (MF$_4$)
takes the hard-en lead at 4.37\%, while FishAudio~S2 and VoxCPM~2
lead the other three monolingual subsets, with \dotstts{}
competitive but not leading. The Pretrain~$\to$~SOAR drop on
hard-en (5.99~$\to$~4.49) is the largest SOAR-stage gain we
observe across this section's benchmarks, and MF$_4$ inherits it
through distillation.

On the cross-lingual subset, \dotstts{} (SOAR) leads SIM in both
directions at 75.0 (en$\to$zh) and 72.8 (zh$\to$en), 6 to 8
absolute points above CosyVoice~3 (66.9 / 66.4). Cross-lingual
WER trails CosyVoice~3 (zh$\to$en 5.66\% vs 4.32\%, en$\to$zh 10.75\%
vs 8.01\%).

\subsubsection{EmergentTTS-Eval}
\label{sec:exp:eval:emergent}

EmergentTTS-Eval uses a Gemini-2.5-Pro-0506 audio
judge to compare systems head-to-head against a fixed
gpt-4o-mini-tts reference, across six expressiveness-oriented
scenarios: Emotions, Foreign Words, Paralinguistics, Complex
Pronunciation, Questions, and Syntactic Complexity. We report both
ASR-based WER and the judge's preference win-rate (higher is better)
on the canonical zero-shot configuration of \dotstts{}. For open-source
systems we use \emph{basic\_ref\_en}\footnote{\url{https://github.com/SWivid/F5-TTS/blob/main/src/f5_tts/infer/examples/basic/basic_ref_en.wav}}
as the common zero-shot voice-clone prompt, with all runs in
continuation voice-cloning mode.

\begin{table}[t]
\centering
\small
\setlength{\tabcolsep}{4pt}
\caption{Selected rows from EmergentTTS-Eval, sorted by overall. $\star$ marks
closed-source / commercial systems. \textbf{Bold} marks the best
per column, \uline{underline} the best open-source per column.
Win-rate is judged head-to-head against gpt-4o-mini-tts by
Gemini-2.5-Pro-0506.}
\label{tab:emergent}
\resizebox{\textwidth}{!}{%
\begin{tabular}{llcccccccc}
\toprule
\textbf{Model} & \textbf{Voice} & \textbf{WER(\%)$\downarrow$} & \textbf{Overall$\uparrow$} & \textbf{Emo.$\uparrow$} & \textbf{Paraling.$\uparrow$} & \textbf{Foreign$\uparrow$} & \textbf{C.\,Pron.$\uparrow$} & \textbf{Quest.$\uparrow$} & \textbf{Syntax$\uparrow$} \\
\midrule
Gemini-2.5-Flash-TTS$^\star$ & Zephyr & 10.39 & \textbf{70.7\%} & \textbf{95.9\%} & \textbf{91.3\%} & 58.5\% & 55.7\% & \textbf{63.0\%} & 57.9\% \\
Gemini-2.5-Pro-TTS$^\star$   & Zephyr & 11.79 & 69.3\% & 86.9\% & 82.3\% & 58.2\% & \textbf{64.8\%} & 61.3\% & 61.8\% \\
gpt-4o-audio-preview$^\star$ & Ballad & 11.87 & 65.2\% & 88.8\% & 82.1\% & \textbf{60.2\%} & 40.4\% & 57.0\% & 59.5\% \\
gpt-4o-mini-tts$^\star$      & Alloy  & 10.76 & 56.3\% & 59.2\% & 58.8\% & 57.3\% & 52.4\% & 52.7\% & 57.1\% \\
\textit{BASELINE: gpt-4o-mini-tts} & Alloy & 10.61 & 50.0\% & --- & --- & --- & --- & --- & --- \\
\rowcolor{rowhl} \textbf{\dotstts{} (Pretrain)} & basic\_ref\_en & 10.86 & \uline{49.2\%} & \uline{72.7\%} & 54.7\% & \uline{39.5\%} & 18.0\% & 48.4\% & 58.4\% \\
\rowcolor{rowhl} \textbf{\dotstts{} (MF4)}      & basic\_ref\_en & 11.75 & 47.9\% & 59.8\% & 55.2\% & 36.3\% & 16.7\% & 50.5\% & 64.8\% \\
\rowcolor{rowhl} \textbf{\dotstts{} (SOAR)}      & basic\_ref\_en & \uline{10.45} & 47.6\% & 63.9\% & 52.7\% & 39.4\% & 16.4\% & 47.0\% & \uline{\textbf{65.7\%}} \\
Qwen3-TTS                    & basic\_ref\_en & 17.32 & 42.8\% & 39.8\% & 50.7\% & 25.4\% & \uline{30.0\%} & 48.9\% & 60.4\% \\
HumeAI$^\star$               & ---    & 12.85 & 42.7\% & 61.6\% & 36.9\% & 34.6\% & 34.3\% & 43.2\% & 44.6\% \\
Qwen3-TTS                    & Ryan   & 19.65 & 42.3\% & 60.5\% & \uline{62.7\%} & 17.1\% &  9.8\% & \uline{56.4\%} & 43.0\% \\
VoxCPM2                      & basic\_ref\_en & 11.84 & 41.1\% & 42.3\% & 44.1\% & 33.3\% & 18.6\% & 53.4\% & 52.3\% \\
MiniMax/speech-02-hd$^\star$         & EN-narr & \textbf{10.02} & 36.6\% & 40.9\% & 34.3\% & 34.3\% & 16.3\% & 47.3\% & 43.9\% \\
11Labs Multilingual~v2$^\star$       & Brian  & 11.19 & 33.9\% & 30.4\% & 45.5\% & 35.5\% & 14.5\% & 39.5\% & 35.5\% \\
F5-TTS                       & basic\_ref\_en & 16.47 & 15.3\% & 26.8\% & 21.6\% & 1.8\% & 1.4\% & 14.8\% & 23.8\% \\
\bottomrule
\end{tabular}%
}
\end{table}

\Cref{tab:emergent} reports the results. \dotstts{} (Pretrain)
leads the open-source field on overall win-rate at 49.2\%, with
SOAR and MF$_4$ close behind (47.6\%, 47.9\%) and all three
variants near the gpt-4o-mini-tts baseline (50\% by
construction). SOAR also has the lowest WER among open-source
systems at 10.45\%.

On Syntactic Complexity, SOAR reaches 65.7\%, the top score
across both open- and closed-source systems (next
Gemini-2.5-Pro 61.8\%). The Pretrain~$\to$~SOAR uplift on this
column is +7.3 points, the largest single-column gain we observe
from the SOAR stage (\Cref{sec:exp:train:sca}). The same shift
drops Emotions (72.7\%\,$\to$\,63.9\%) and Paralinguistics
(54.7\%\,$\to$\,52.7\%), suggesting SOAR tightens text faithfulness
at the cost of expressiveness. On Emotions, Pretrain leads the
open-source field at 72.7\%. Closed-source systems on this
column (87--96\%) use carefully tuned built-in voices, while
\dotstts{} runs zero-shot voice cloning from a basic reference
clip.

On the remaining columns \dotstts{} is mid-field. The weakest are
Complex Pronunciation (16--18\%) and Foreign Words (36--40\%),
both stressing rare or out-of-distribution lexical items.

\subsection{Efficiency}
\label{sec:efficiency}

We report inference-time efficiency on the Seed-TTS-Eval
test set under vllm-omni~\citep{vllmomni}.
The LLM runs on vLLM with continuous batching and paged-KV attention,
while the AR-FM head and the semantic encoder are JIT-compiled with
\emph{torch.compile}. \emph{Time to first-packet latency} (TTFP, ms) is the
wall-clock latency from request arrival to the first emitted audio
packet, and \emph{RTF} is the ratio of generation wall-clock time to
synthesized-audio duration.

All numbers are measured on a single NVIDIA H800 GPU using the
MeanFlow-distilled student (\Cref{sec:exp:train:meanflow}), with the
AR-FM head running at $\mathrm{NFE}=4$ per audio patch. In text-only
synthesis, \dotstts{} runs comfortably in real time, reaching RTF~0.231
in plain mode and RTF~0.245 in 1T1A interleaved mode. Interleaving
reduces first-packet latency from 85.4~ms to 54.4~ms by consuming the
upstream LLM token stream as it is decoded, so audio generation can
start once the large language model begins emitting its response. 
\section{Conclusion}
\label{sec:conclusion}

We presented \dotstts{}, a 2B-parameter fully continuous, end-to-end
autoregressive TTS system that targets the two open problems we
identified for the continuous-AR paradigm: long-range error
accumulation during AR rollouts, and an immature post-training stack
relative to the discrete-token cascade. The backbone is decomposed
into a semantic encoder $\to$ LLM $\to$ autoregressive flow-matching
head, with the audio-side input to the LLM restricted to a 6.25~Hz
semantic summary of each newly generated patch, not the raw VAE
latent. This decoupling keeps the LLM operating on a compact semantic
view of the history, important for long-rollout stability. On the
synthesis side, a high-fidelity continuous AudioVAE (trained with a
WavLM-alignment loss and a multitask downstream block so that the
high-rate continuous latent remains learnable for the downstream LLM)
preserves the timbral and paralinguistic detail that low-bitrate
discrete codecs typically flatten. We complement the pretraining recipe with Self-corrective
alignment, a reward-free, flow-matching-native post-training stage that
teaches the acoustic DiT to recover from its own inference-time errors,
and with CFG-aware MeanFlow distillation, which enables few-step
generation with a single conditional model evaluation per step.

Trained on 1.5M hours of speech, \dotstts{} attains
state-of-the-art average WER (2.92) and SIM (79.2) on
Seed-TTS-Eval, the highest average speaker
similarity (83.9) on the 24-language MiniMax multilingual
benchmark, and leading hard-subset and
cross-lingual results on CV3-Eval. On
EmergentTTS-Eval, it takes the top Syntactic
Complexity score in the table (65.7\%, ahead of every closed-source
system) and is the strongest open-source system on Emotions (72.7\%). Combined with CFG-aware MeanFlow
distillation and a 1-text-1-audio interleaved streaming layout, the same
backbone reaches 54~ms TTFB at RTF~0.245 on a single H800, making it
deployable for real-time and conversational use cases. We release the
full training and inference code, together with the pretrained,
self-corrective-aligned, and MeanFlow-distilled checkpoints, under the
Apache~2.0 license as a reproducible reference stack for continuous-AR
TTS.

\section{Limitations}
\label{sec:limitations}

Several limitations remain in the current system. Feeding the LLM raw BPE rather
than phonemes inherits its text capabilities at the cost of higher data appetite,
which on the script-divergent and under-represented languages of
\Cref{sec:exp:eval:minimax} (Arabic, Hindi, Turkish, Vietnamese) drives
the low-resource WER gap visible there, and on the Foreign Words
and Complex Pronunciation scenarios of \Cref{sec:exp:eval:emergent} hits the
same coverage limit on loanwords, technical terms, and proper names.
Expanding the multilingual pretraining mix on those languages, adding a small
phoneme-side auxiliary input, and inserting a language-balanced post-training
stage are direct levers to address this.
The released system is also evaluated only under canonical
zero-shot conditions, without explicit style or instruction control. An
instruction-tuned variant built on the caption-paired data of
\Cref{sec:exp:data} is a natural next step. Although the AudioVAE is
in principle modality-agnostic, the backbone is trained
on a speech-heavy mixture, so singing and unified speech-and-sound generation are
not covered in this release. Finally,
high-fidelity zero-shot voice cloning carries well-known misuse risks. The
released checkpoints are intended for research and authorized deployment, and we
encourage downstream users to combine them with consent-aware reference-audio
policies, robust synthetic-speech detection, and content watermarking.

\section*{Contributors}
\label{sec:contributors}

\dotstts{} is jointly developed by dots, Xiaohongshu
Inc.\textsuperscript{1} and the X-LANCE
Lab\textsuperscript{2} at the School of Computer Science, Shanghai Jiao
Tong University. Both teams contributed to the model design, the
training recipe, and the evaluation, and the checkpoints, code, and
this report are released jointly.

\paragraph{Authors}
Shi Lian\textsuperscript{1},
Changtao Li\textsuperscript{1},
Bohan Li\textsuperscript{2},
Hankun Wang\textsuperscript{2},
Da Zheng\textsuperscript{1},
Junfeng Tian\textsuperscript{1},
Yufeng Ma\textsuperscript{1},
Colin Zhang\textsuperscript{1},
Kai Yu\textsuperscript{2}.

\bibliography{references}

\end{document}